\documentstyle[12pt,amssymb,epsfig]{article}
\begin{document}

\baselineskip18pt

\begin{center}
{\Large\bf Population synthesis in astrophysics}

\vspace{1cm}

{S.B.~Popov$^1$, and M.E.~Prokhorov$^{1}$}

\vspace{0.5cm}

$^1$ Sternberg Astronomical Institute, Universitetski pr. 13,
119992, Moscow, Russia, 
polar@sai.msu.ru, mike@sai.msu.ru \\
\end{center}

\vspace{1cm}

\begin{abstract}
We briefly describe different astrophysical applications of a
population synthesis method.
In some details we discuss the population synthesis of close binary systems
and of isolated neutron stars.
\end{abstract}

\section{Introduction}

 In this lecture notes we give a brief introduction into 
 {\it population synthesis in astrophysics}.  

A {\it population synthesis}  is a method of a direct modeling of relatively
large populations of weakly interacting objects with
non-trivial evolution. As a rule, the evolution of the objects is followed from
their birth up to the present moment.
  
In the following sections we describe a necessity
of this type of modeling and give a general description of the method
(secs. 2 and 3).
Then in some details we discuss studies of close binaries (sec. 4) and
of isolated neutron stars (sec. 5).  In the sixth section we give examples
of other types of population synthesis studies; namely, 
we describe  modeling of galactic
spectra via simulations of stellar populations and  modeling of the X-ray
background radiation. Finally we present our conclusions. 
Since  many examples are just briefly
touched, we provide a list of references to important papers and reviews.

 Before we start it is useful to make one comment on our definitions.
There are two main approaches and both are sometimes called {\it
population synthesis}. One has a very wide area of application. In this
method calculations are started with some initial conditions, and a
population is followed from its birth. Some authors call it {\it
evolutionary synthesis} (see \cite{Fritze-v.Alvensleben00} 
for example). However, more often it is
called {\it evolutionary population synthesis} (see for example
\cite{CidFernandes&01,Maraston04,Bruzual&Charlot03}). 
Another method, used mainly for studies of stellar populations (see
sec.6.1),
is simply called {\it population synthesis} in \cite{Fritze-v.Alvensleben00} 
and in some other papers. 
This method is often called {\it empirical population synthesis}
(see \cite{CidFernandes&01}).
The second way of notation is probably better, as it allows not to mix the two
approaches but to save the main part of notation
-- population synthesis -- in both.
Below we shall mainly speak about the first method (except sec. 6). 
If it does not mislead the reader, we shall  call it just {\it population 
synthesis} (PS).


\section{Necessity of population synthesis studies in astrophysics}

 A necessity of the PS in astrophysics is mainly
determined by a specific character of astrophysics as a science.
We cannot make direct experiments with objects under study: experiments
are replaced by observations. An observation in a sense is just  a snapshot
of some object(s) with very long evolutionary time scale. It is impossible
to follow
an evolution of a given galaxy or a star simply observing it. Often an
evolutionary path can be explored  only by modeling.

Another specific feature of astronomy is that {\it ``we are searching where
it is brighter''}. I.e. sometimes we observe objects which are not typical
representatives of some population, but oppositely are very peculiar.
We can make an attempt to construct a theory to describe a few particular
observed sources of some type. Nevertheless, it is necessary to check if this
theory is consistent with an existence of other (probably unobserved yet)
sources of the same type. Thus, finally it is necessary to study populations,
not just single (easily observable) objects. 

It is important to note that in many
cases observations of individual objects are not available, only integral
characteristics can be obtained. In that case for better understanding of
properties of an observed population it is also necessary to model it. 

 The third feature we want to note is like that.
Many astronomical projects (satellites, networks of telescopes, etc.) are
very expensive. It is necessary to justify their construction in advance. 
This means that it is necessary, for example, 
to convince a community that new interesting,
previously unobserved objects can be found. It demands an analysis of
populations of unobserved sources basing on any available data on their
already detected ``relatives''.

 Two main goals of a population synthesis modeling are the following.
At first the PS can be used to test or derive
some parameters of a population under 
consideration (for example, initial spin periods of neutron stars, an initial
mass function of stars) by comparison of a real (observed) 
and a theoretical (modeled) populations of sources. 
In astronomy very often
we cannot directly obtain initial and evolutionary
parameters from observations in a model independent way. 
The PS is one of very few instruments to derive unknown properties
of astronomical objects via a computer modeling. If the whole set of parameters 
of a population
is successfully tested, i.e. an output of a model is in good correspondence
with observations, then we have a quantitative explanation of the origin and
the evolution of the discussed type of objects. 

The {\it Absolute PS}
of that kind might contain an artificial universe in a computer. 
If all properties of the artificial and real universes coincide, we 
can conclude that we have build a self-consistent picture of the physical
world with all initial and evolutionary parameters in correspondence with 
data of observations and experiments.  
So we can call the first approach as {\it testing and explanation}. 

 The second goal of a PS model is to {\it predict} properties of unobserved
sources (for example, the coalescence rate of binary supermassive black holes
detectable by LISA, or the rate of neutron stars coalescence detectable by
LIGO). In that case uncertainties of a prediction are mainly
determined by uncertainties of the input parameters. 
{\it ``The reliability of a population synthesis is as good as the input
assumptions''} \cite{Verbunt&99}. 
Of course, a future comparison with new observational
data will give an opportunity to test assumptions of a model.

 Now let us discuss some basic features of a PS 
which may be applied to models of different kind.


\section{General properties of population synthesis models}

In our opinion the PS method as it is used in astrophysics does not have
direct analogues in other fields of science. 
For example, most of objects studied
in physics (elementary particles, atoms, molecules, etc.) are much simpler
than astronomical objects and have much more trivial evolution. On the
other hand, behaviour of living creatures is much more complicated;
interactions between different individuals and species might be taken into
account.

All parts of the definition of the PS presented in the Introduction are
equally important. 

\noindent
{\it i).} The population under consideration might not be very small.
Otherwise if calculations are made for the same (small) number of objects,
statistical fluctuations will dominate in the final results.
If oppositely calculations are made for a large number of objects, then
a comparison with a small actual population is not informative.
A similar problem can appear even for relatively large populations if a few
rare objects can dominate in the final integral value of some parameter.
For example a few rare luminous stars (or even one star) can dominate in the
integral colour of a cluster 
(see~\cite{Cervino&Luridiana03,Cervino&Luridiana04} for a discussion of some
limitations of the PS method).

\noindent
{\it ii).} An evolutionary path of each  object might not be very simple
or very complicated. In the first case a PS itself becomes unnecessary,
as analytical calculations can be used (however, 
in early PSs of single and binary
stars in the frame of very simplified models authors applied analytical
calculations). If an evolutionary track is too complex then it is just
difficult to model it, and in addition the number of possible states becomes
too large, possibly exceeding the number of objects in the observed population.   

\noindent
{\it iii).} Objects should not significantly interact with each other,
otherwise it is impossible to follow individual tracks. However,
interactions
are not absolutely forbidden (for example, stars in globular cluster can
interact and form or destroy binaries).
  
From the philosophical point of view it is possible to divide most of PS
calculations into two stages:

\begin{itemize}\item Construction of a model population of objects.
\item Calculation of quantities of interest (and their distributions)
using the constructed population.
\end{itemize}

Quantities of interest can include: a number of particular sources, 
average or extreme
values of their parameters and statistical moments of these parameters,
correlations between given pairs of parameters, different parameter
distributions which can be one-parametric or multi-dimensional, 
cumulative or in the differential form, etc.

If a set of sources itself (the whole population or its part) is not the
main goal of a study, then the set may be constructed implicitly. 
In such a case
a task of calculating of 
a parameter or its distribution is formally reduced to
a multi-dimensional integration of a complicated function which on its part
depends on an evolutionary scenario. To calculate such integrals one can use
multi-dimensional regular grids or the Monte-Carlo method. Each of this two
approaches has advantages and drawbacks.

 The way of integration (the classical one or the Monte-Carlo) depends on
a method of setting initial parameters of the objects under consideration.
A chosen range of initial parameters should cover all the region of
initial conditions which is valid for a given problem. Then the coverage
should be uniform enough. 
If an integration proceeds in the classical way, then a
chosen region of initial conditions should be covered by a
(quasi)rectangular grid, not necessarily uniform. In the
Monte-Carlo method initial conditions are chosen stochastically with
statistical weights proportional to an assumed actual realization. The
second method provides the most realistic populations of modeled objects.


\section{Population synthesis of close binary systems}

In this section we describe the PS of binary stars. 
In particular we shall focus
on systems with compact objects (neutron stars or black holes). 
A detailed description of the evolutionary scenario for binary systems
can be found in~\cite{LPP96}.\footnote{This review can be found on the Web at
http://xray.sai.msu.ru/$\sim$mystery/articles/review/ .}
We start with a very brief description of the stellar evolution (details can
be found in many standard textbooks, see for example
\cite{Bisnovatyi-Kogan02} and references therein). 

It is possible to distinguish two main parts of a star's life: when it is a
normal non-degenerate star, and when it becomes a compact stellar remnant
(a white dwarf, a neutron star or a black hole).
Thermonuclear reactions are an energy source of normal stars (see lecture
notes by D.K. Nadezhin in this volume). In some sense an evolution of a normal
star can be reduced to changes of the nuclear fuel 
(central hydrogen, hydrogen in a thin layer, helium, etc. 
up to iron for the most massive stars).
About 90\% of its life a normal star spends at the main sequence burning
hydrogen in its a center. After the main sequence stage 
a star can significantly
change its radius, surface temperature and loose large part of its
mass (mass losses at the main sequence are important only for very massive
stars).  Finally a star becomes a compact remnant. Objects with initial
masses $\lesssim8$-$10 \, M_{\odot}$ become white dwarfs (of course for
stars with $M\lesssim 0.8 \, M_{\odot}$ an evolution prior to a white dwarf
formation can be very long, longer than the present age of the Universe).
More massive stars explode and form neutron stars or black holes. The
boundary which divides neutron star progenitors from black hole progenitors
is uncertain. Roughly this value is about 25--35~$M_{\odot}$. On the other
hand,
there are suggestions that it can depend on rotation and magnetic field
of a star (see \cite{Woosley&02} 
for a detailed discussion of late stages of the stellar evolution).   

 The main parameter governing the evolution of a single normal star 
(and in some sense of a single compact object) is its mass. An isolated star
can only loose its mass due to a stellar wind. In a binary the situation is
drastically different: a star may become a donor or an acceptor, i.e. it can
transfer matter to a second companion or to get mass from it. This option
immediately shows that the evolution in a binary is much more complicated.

 A small amount of mass can be transferred via a stellar wind, but the main
channel of a mass transfer
is an accretion due to the Roche lobe overflow. For example, in such a process, 
a normal massive star can transfer all its outer layers and become
a Wolf-Rayet star -- a naked helium core. If the secondary star is a compact
object, then an accretion results in an X-ray source formation. 

 Different types of compact objects have distinct evolutionary paths.
A black hole can increase its mass due to an accretion in a binary. 
Because of the same reason it can be also spun-up. 
Isolated black holes can be called nearly non-evolving objects.
For isolated white dwarfs the most important aspect of evolution is cooling.
The richest life full of different stages is a fate of neutron stars,
especially in binaries. 

A neutron star can appear as a visible source due to its thermal emission,
due to the radio pulsar activity, 
due to an accretion onto its surface, or due to
an interaction of its magnetic field with a surrounding plasma. 
A neutron star's surface temperature, its spin period and magnetic field (plus 
parameters of the
surrounding medium) are the main parameters which 
determine an astrophysical appearance of a neutron star.
Therefore it is possible to distinguish 
at least two  types of evolution of these
objects: thermal history and magneto-rotational evolution.  
For some sources these two types
of evolution do not influence each other, for some they do (for example for
magnetars).


The first PS studies of binary stars were 
performed at the beginning of 80s
\cite{Kornilov&Lipunov83,Lipunov&Postnov87,Dewey&Cordes87,Manteiga&89,Yungelson&Tutukov91}.
These authors utilized the evolutionary scenario constructed 
by Paczy\'nski, Kippenhahn,
Weigert, Iben and many others (see reviews in \cite{Paczynsi71,vdHeuvel94}).

Here we shall discuss the PS of close binary systems with compact objects
(white dwarfs, neutron stars and/or black holes). The evolution of binary
stars produces a variety of different objects with different observational
manifestations. In this notes we cannot describe all aspects of the problem,
so we just shortly mention some types of objects and some observed
phenomena.
We shall discuss
cataclysmic variables, binary white dwarfs, 
X-ray binaries, millisecond radio pulsars, and ultraluminous sources. We
describe results of researches of the following phenomena: a common envelope
stage and type Ia supernovae. Then we present main results and difficulties in 
studies of gravitational wave sources. We discuss coalescence of neutron stars 
and black holes
and gravitation radiation from binaries. In the last part of the section we
discuss binary evolution in dense stellar clusters taking into consideration
far and close stellar encounters.

Before we start to discuss different types of sources and phenomena 
let us give a short (and probably incomplete)
list of active scientific groups working on the PS
of close binaries and references to some of their recent results in order to
provide a guide for the readers interested in this subject:

\begin{itemize}

\item
Portegies Zwart et al.: 
PS of dense stellar clusters \cite{PZ&01,Pooley&03,Sills&03,PZ&04b}, 
intermediate mass black holes \cite{PZ&04a,PZ05}, 
X-ray sources \cite{Brown&01,Nelemans&04}, etc.

\item
Podsiadlowski et al.:
evolution of compact binaries and neutron stars kicks 
\cite{Pfahl&01,Pfahl&02,Pods&04a},
hypernovae and gamma-ray bursts \cite{Pods&95,Pods&04b},
other questions~\cite{Beer&Pods02,Han&Pods04}.

\item
Kalogera et al.:
neutron stars coalescence rate and gravitational radiation
\cite{Kalogera&Lorimer00,Kalogera&Belczynski01,Kim&03,Kalogera&04},
radio pulsars, neutron stars and black holes
\cite{Fryer&Kalogera01,Willems&Kalogera04}.

\item
Yungelson et al.:
double white dwarfs and supernova type Ia
\cite{Yungelson&Livio00,Fedorova&04,Napiwotzki&04,Yungelson04},
stellar gravitational wave sources
\cite{Nelemans&01}, etc.

\end{itemize}

\subsection{Census and distributions of compact stars}

The PS of binaries is a useful tool to study statistical
properties of stars including compact objects. We wish to address the
following questions: 

\begin{itemize}
\item How many compact objects of different types are born? 
\item What is the mass distribution  of these compact objects? 
\item What is the relation between a stellar initial mass 
and a final mass of a compact object 
taking into account the binary evolution? 
\item With a given distribution of compact object
masses, what are relative numbers of different types of objects (neutron
stars, quark stars, black holes)?
\end{itemize}

The binary PS is a hot point of astrophysics. Different scientific groups all 
over the world have there own codes for calculations of the binary evolution, 
and produce a huge flow of publications on different aspects, discussing all the 
questions mentioned above
(see, for example,
\cite{Pols&Marinus94,Bethe&Brown98,PZ&Yungelson98,Bloom&99,Belczynski&Bulik99,Belczynski&02,Tutukov&Yungelson02}).
Below we briefly mention some of these results referring to original papers
for details. 

Pols and Marinus~\cite{Pols&Marinus94} calculate an
evolution of young open clusters. The authors focus on blue
stragglers. There are several hypothesis on the nature of these sources.
They can be binaries consisting of two main
sequence star or one main sequence star and a compact object; or they can be
single pecular stars, for example formed by merging of two normal stars. 
Distributions of age and other parameters are presented.

In \cite{Bethe&Brown98} the authors consider merging binaries consisting of
two neutron stars or black holes. An estimate of a number of binary radio
pulsars (systems like the Hulse-Taylor radio pulsar) is made.

Ref. \cite{PZ&Yungelson98} is devoted to formation and evolution of binary
neutron stars. The amount of young and recycled binary
pulsars is calculated for different assumptions on kicks, 
initial mass ratio, and presence or absence
of hyperaccretion regime.

The main goal of \cite{Bloom&99} is to study coalescence of binary
neutron stars with implications to gamma-ray bursts (now this scenario is
mainly applied to short gamma-ray bursts, while long bursts are explained as
hypernova events). The authors estimate birth and merging rates, spatial
velocities; also they calculate spatial evolution
of the binaries and their displacement from the host galaxies.

In \cite{Belczynski&02} the mass distribution of compact objects is
calculated for single and binary stars for different types of
a secondary component. The authors include into their model
a possibility of a quark star formation. However, their results are very
model dependent, and in general the mass spectrum of compact objects is
poorly known now (see also sec 5.3 below).

At the end of this subsection 
we want to attract your attention to two important notices. 
First, simulations
considered above use different evolutionary tracks of normal stars
(analytical approximations, evolutionary track libraries, etc.), various
initial distributions (initial mass ratio, kicks, magnetic fields, etc.), all
sorts of assumptions about complex stages of the evolution (common envelope,
accretion induced collapse, merging, etc.). Second, in the final
results of different calculations there are some discrepancies, 
though general predictions are very similar and robust.

\subsection{Cataclysmic variables}

A cataclysmic variable is a binary system consisting of an accreting white dwarf 
and a low-mass main sequence
star filling its Roche lobe. 
The state with a Roche lobe overflow is supported by angular momentum losses
due to the magnetic stellar wind braking or due to gravitational radiation. 
Magnetic wind originates due to a 
presence of a convective envelope of a normal star. This mechanism works 
for stars in the mass range
0.3-1.5~$M_\odot$. 
When the mass of a donor star becomes smaller than $\sim 0.3\, M_{\odot}$
the magnetic braking vanishes and accretion stops. Later
the evolution of the binary is
determined by the gravitational wave radiation. 
After a period of an X-ray quiescence the 
normal star fills its Roche lobe again, and the X-ray source switches on.
The following evolution of such object is mainly driven by emission of the
gravitational waves.

The magnetic braking accretion is more powerful than accretion driven by
gravitational wave momentum losses, and
the corresponding sources are brighter. 
Besides the magnetic braking and gravitational waves there is another way
to loose orbital momentum: an interaction with a circumbinary disc. 
Ordinarily this mechanism is  less effective.

The model described above predicts a gap in the orbital period distribution. 
This is really observed at $P_{\mathrm{orb}} \sim 3$ hours. 
Almost all long period cataclysmic variables are induced by 
the magnetic stellar wind, all short
period ones ($P_{\mathrm{orb}}<3$~hour) are gravitational wave induced. 

A PS study of cataclysmic variables and their
orbital period distribution in a classical approach
was performed in many publications
(see for example~\cite{Kolb98,Kolb&98,Howell&01}).
The authors are able to reproduce the observed orbital distribution
of the variables.
The PS of cataclysmic variables, surrounded by a circumbinary disc
is performed in 
\cite{Willems&Kolb04,Kolb&Willems04}. 
The authors come to the conclusion that
the bimodal period distribution in the frame of
the outer disc model is possible only if a proper bimodality
of disc feed properties is presented.

\subsection{Binary white dwarfs}

In applications of a PS to binary white dwarfs there are two topics
which seems to us the most interesting.

At first, the amount of very close binary white dwarf systems 
can help to constrain the efficiency of angular momentum losses 
at the common envelope stage. Most of the close
binary white dwarf systems are formed after two common envelope episodes.
The first one is due to the Roche lobe filling by the primary,
and the second starts after the secondary component also fills its Roche
lobe. 
This type of systems is the most sensitive indicator of 
common envelope parameters. According
to~\cite{Iben&97,Nelemans&Tout04} the parameter of the efficiency is 
$\alpha_{\mathrm{CE}}\simeq0.3$--0.5.
This value does not coincide with results of other PSs calculations which take 
into
account evolution with just a single common envelope stage~\cite{Politano01}.

Coalescence of binary white dwarfs is the second important topic.
This process is one of
evolutionary channels leading to a type Ia supernova explosion -- so-called
double-degenerate scenario.\footnote{The second possibility to produce such 
supernova is the
accretion induced collapse of a white dwarf -- the single-degenerate
scenario.} This topic was studied, for example, 
in~\cite{Kalogera&Lorimer00,Kalogera&Belczynski01,Nelemans&01b,Kalogera&04,%
Yungelson04,Kim&03}. It is shown that coalescence of white dwarfs is 
able to produce type Ia supernovae with a rate $\sim$$10^{-3}$~yr$^{-1}$.
This estimate is valid for both old and young stellar populations
(i.e. for elliptical and spiral galaxies). 
For the second channel (accretion onto a white dwarf from a normal
companion) calculations give an
estimate of the supernova Ia rate $\sim$$10^{-4}$~yr$^{-1}$.  
However, it is worth noting that recent studies show that the supernova Ia
rate per unit mass is much higher for young stellar populations
\cite{Manucci04}. This can be an argument against the double-degenerate
scenario.

\subsection{Accreting X-ray sources and millisecond pulsars}

Present day orbital X-ray observatories like  Chandra and XMM have a
sufficient sensitivity to detect all bright binary X-ray sources in the
Galaxy and almost all sources with moderate luminosities. There are many
different types of binary X-ray sources with permanent or transient hard
emission: X-ray pulsars, X-ray bursters, so-called ``atoll'' and Z-sources,
sources with QPO, X-ray novae, black hole candidates etc. Some sources may
display several of the enumerated phenomena simultaneously. Observational
properties of different types of the sources are discussed in the following
reviews~\cite{Joss&Rappaport84,Verbunt93,Tanaka&Shibazaki96,vdKliss00,Swank&Markwardt01,Grimm&02}.

The first PS investigations of high-mass X-ray binaries were made 
about 20 years ago. 
PS studies of low and intermediate-mass X-ray binaries with more complex
evolution started later. 
Recently different authors started to include into calculations individual
properties of studied regions; 
for example, starformation history in a given galaxy, 
or mass and metallicity of a given region of starformation.
Below we give a short list of recent publications devoted to PS of binary X-ray
sources. 

In the papers \cite{Dalton&Sarazin95,Terman&98,Iben&95} 
and in a series of papers \cite{Iben&95} 
the authors calculate accretion rates, luminosities, spatial
velocities, and birth rates of massive X-ray binary sources with supergiants
and Wolf-Rayet companions. In~\cite{Terman&98} a population of X-ray sources
with Be-companions is studied. 
In \cite{Dalton&Sarazin95} --
a population of the Thorne-Zytkow objects 
(red giants with neutron star core --
a possible result of merging of components of a
massive binary during a common envelope stage).
In~\cite{Lipunov&96} a population of X-ray sources near the center of 
our Galaxy is synthesized. 
Comparison of observed luminosity distributions of
X-ray sources with PS results of \cite{Kinwah01} indicates the presence of
several populations of X-ray binaries in nearby galaxies. Each
population has its own formation and evolutionary history, 
depending on the host environment.

In~\cite{Podsiadlowski&01} the authors 
try to reproduce about 140 observed low-mass
binary galactic X-ray sources in the frame of low and intermediate-mass binary
evolutionary scenario. Ref. \cite{Pfahl&03} 
represents a  more advanced version of the previous PS by the same team. 
In this publication properties of millisecond pulsars are also considered.

In the end of this subsection
we want to emphasize two interesting, 
in our opinion, groups of X-ray sources.

Various X-ray systems are different not only in a physical sense, 
but also in PS contents. 
Equal observational amounts of sources may be produced by
numerous short lived objects or by rare long lived ones. A PS of the latter 
objects is a sufficiently more difficult task. The most interesting
representatives of that sort of objects are accreting progenitors of
millisecond radio pulsars. A PS of millisecond pulsars and their progenitors
is discussed in~\cite{Possenti&98,Willems&Kolb02}.

X-ray novae are accreting close binaries consisting of a
black hole and a low mass main sequence component. 
To create such a system it is necessary to start with a 
binary with very high initial mass ratio.  A black hole progenitor have to
be at least as massive as 25-$30\,M_\odot$. Of course such a massive star
looses significant part of its mass prior to a compact object formation.
A low-mass component saves its initial small mass.
It is not easy to reproduce a significant number of X-novae in the frame of
a standard scenario of binary evolution. In 1986 Eggleton and
Verbunt~\cite{Eggleton&Verbunt86} suggested a scenario for 
creation of an X-ray nova from a massive hierarchical triple system. 
The evolutionary scenario for
hierarchical triple and multiple stellar systems posses sufficiently more
potential variety instead of the binary evolution scenario (see, for
example,~\cite{Mardling&Aarseth99}). Computer realization of an evolution
of a triple system is very complex,
At present we know only one realization of such code. 
Comparison of these two scenarios for the X-ray nova evolution 
and results of binary and triple PS calculations  can be found
in~\cite{Kuranov&01}.

\subsection{Ultraluminous X-Ray sources}

Ultraluminous X-ray sources are very bright point X-ray objects outside 
nuclear regions of host galaxies 
(see a catalogue in \cite{Colbert&Ptak02} 
and a recent review in \cite{Mushotzky04}). 
Ordinarily this objects are defined as sources with 
an X-ray luminosity
larger than $10^{39}$~erg~s$^{-1}$ (from $\sim 10^{39}$ up to  $\sim 
10^{42}$~erg~s$^{-1}$).
They were discovered by Einstein space observatory \cite{Fabbiano89}. 
Large number of these sources was later 
found by ROSAT and Chandra. In our Galaxy bright ultraluminous sources are 
absent.

There are two popular models of ultraluminous sources.
In one of them an accretion with nearly  Eddington rate proceeds onto an
intermediate mass black hole ($10^2$--$10^3\,M_{\odot}$) in a binary system
(e.g. \cite{Colbert&Mushotzky99}, see \cite{Miller&Colbert04} for a detailed
discussion on the properties of intermediate mass black holes). 
In another hypothesis it is supposed that we deal with normal stellar mass
black holes, but radiation is not symmetric, being beamed into two jets.
If we look along the jet, then we observe a very bright source.
In this model ultraluminous sources may be just a high luminosity tail of the
stellar-mass black hole luminosity distribution 
(e.g. \cite{King&01,Postnov03}).
Both hypothesis have advantages and drawbacks. 
Probably the ultraluminous sources population is formed by two (or even more?) 
different types of objects~\cite{Soria&04}.

Podsiadlowski at al. \cite{Podsiadlowski&03} perform a systematic study of
formation and evolution of black hole binaries and find that indeed
their models are consistent with the observed luminosity function, and a
typical number of ultraluminous sources per galaxy can be naturally
explained.
Note, that 
as the ultraluminous sources are very rare, bright sources of that type are
observed with a rate 0.01 per galaxy, calculations of their luminosity
function and its comparison with observations is not trivial. Different
statistical and selection effects might be taken into account.

\subsection{Stellar gravitational wave background}

New generations of ground-based (LIGO, VIRGO, GEO) and space (LISA)
gravitational wave interferometric detectors will be sensitive 
at $10^1$--$10^3$ and at $10^{-4}$--$10^{-1}$~Hz, respectively. In this
frequency bands there are three general types of stellar gravitational wave
sources \footnote{We do not discuss coalescence of supermassive black holes, 
which can be an important type of sources for LISA.}:
 
\begin{enumerate}
\item Core-collapsed supernovae (type II, Ib or Ic); 
\item Rapidly rotating non-axisymmeric neutron stars; 
\item Close binaries and coalescing compact stars. 
\end{enumerate}

Let us briefly discuss these items (coalescence of compact objects is also
a subject of the next subsection).
The rate of supernovae is roughly  known ($\sim$1/30-1/50~yrs$^{-1}$).
Nevertheless, to calculate a possible rate of detection of gravitational signals
it is necessary to know how much energy is transferred into gravitational
radiation during an explosion.
As a complete self-consistent theory of supernovae is absent,
the fraction of the energy of an explosion which goes into gravitational
waves is unknown. This is the most important uncertainty of this subject. 

Rapidly rotating neutron stars emit monochromatic gravitational waves.
The energy flux may depend on a proper non-sphericity of a neutron star.
For very fast neutron stars the symmetry breaking appears  due to the
Chandrasekhar-Friedman-Schutz instability
\cite{Chandrasekhar70,Friedman&Schutz78}, or due to an r-mode instability
\cite{Kokkotas&Stergioulas99,Andersson&99}.
In this case the flux depends on viscosity and temperature of
a neutron star crust and interior.
All parameters here, especially viscosity,  are not well known. Results of 
different studies are
still very controversial.

The gravitational radiation from binary stars is well understood 
and inevitable
in the frame of the General Relativity. The first estimate of
the gravitational wave
background from galactic binaries 
was obtained by
Mironovcskij in 1965~\cite{Mironovcskij65}  on the base of counts of W~UMa 
systems. This author got an estimate of the dimensionless amplitude of
metric perturbations equal to $h\sim10^{-21}$.
The spectrum of the
stellar galactic gravitational wave background was calculated with the 
binary PS model
in 1987~\cite{LPP87gw}. The first calculations show a presence of two
different parts in the spectrum. The first one, 
at low and intermediate frequencies with
a maximum at
$\nu\sim10^{-4}$--$10^{-3}$~Hz, is linked with the gravitational radiation of 
binaries. 
The second represents a high frequency tail of the spectrum.
It appears due to merging of white dwarfs 
and neutron stars. Resent results on this subject can be found
in~\cite{Grishchuk&01,Nelemans03,Cooray04}.

\subsection{Merging and coalescence rate of neutron stars and black holes}

Components of a relativistic binary system 
with an orbital period less than $\sim 14$ hours can
spiral-in and merge in the Hubble time due to the gravitation wave emission.
Coalescence of neutron stars was widely discussed 
as a possible source of gamma-ray bursts
and as a source of strong gravitational wave pulses detectable by
ground-based detectors.

The rate of
neutron stars coalescence is a key parameter in modeling of an observable
rate of gravitational wave bursts. There are two different estimates of this
value: one is based on PS calculations, and another is  based on the number
of observed binary radio pulsars with short orbital periods (like the famous
Hulse-Taylor system). During last twenty years there was some contradiction
between these two approaches. Since the first paper \cite{LPP87gw} and later
on (see \cite{Jorgensen&95} and references therein) PS
models predict the coalescence rate about $10^{-4}$~yr$^{-1}$ for a spiral
galaxy similar to the Milky Way. The direct estimate based on the binary radio
pulsar statistics predict significantly lower coalescence rate 
$\sim 10^{-6}$~yr$^{-1}$~\cite{Phinney91}. 
Later this value has been increased up to 
$2\cdot10^{-5}$~yr$^{-1}$ (for example~\cite{vdHeuvel&Lorimer96,Kim&03}),
nevertheless, the difference did not disappear. 
This method takes into account only binaries where al least one component is
observed as a radio pulsar. In a PS model all merging neutron stars are
taken into account, including radioquiet ones. This difference can partially
explain the lack of convergence between the two estimates. The discovery of
the double binary radio pulsar J0737-3039 with a very short orbital period
\cite{Burgay&03} increases the observational prediction by nearly an order
on magnitude, and so now the problem is almost solved~\cite{Kalogera&04b}. 
The history of the question is briefly described in~\cite{Lipunov04}.
One can follow the references given in that paper.

If one wants to predict a detection rate of gravitational wave signals from
coalescing compact objects it is not enough to know the rate of this events.
It is also necessary to take into account the power
distribution of the sources, i.e. their luminosities in gravitational waves.
For a fixed sensitivity of a gravitational wave antenna the maximum distance
from which we still
can register a coalescence depends first of all on a mass of a
merging binary (more precisely on the so called ``chirp mass'' -- some
combination of masses of the binary components). If ground-based
interferometers can reach their nominal sensitivity they can detect neutron
star merging from a few dozen megaparsec. The coalescence rate of binary
black holes is probably lower by 1-2 order of magnitude in comparison with
neutron star systems, but note that black holes are more massive (at least
by a factor a few). Due to strong dependence of the maximum detection distance
on a binary's mass we can detect an event from sufficiently larger volume of
space. This factor can be even more important and the \emph{detection rate}
of black holes coalescence can exceed the same quantity for binary neutron
stars~\cite{LPP96}. I.e. the first detected signals can be due to
coalescence of a black hole with a neutron star or with another black hole! 

Note, that values of coalescence rates are very sensitive to
details of evolutionary scenario. For example, as progenitors of black holes
are very massive stars they should form relatively wide binaries. 
If an orbit of a
system is not changed during a black hole formation then the time prior the
components merge is very long. Thus in this case the detection rate is
 zero. To change the situation it is necessary to introduce a natal
kick for black holes~\cite{LPP96key}. These is quite possible if a black hole is 
formed
in a two-stage collapse. Recent studies of black hole binaries support the
non-zero value of the kick velocity~\cite{Willems&04}.

\subsection{Binaries in dense stellar clusters}

Evolution of binary populations in dense stellar clusters (globular clusters
and central, i.e. circumnuclear, clusters in~   galaxies) 
differs from evolution of the same binaries
in the galactic field. There are two reasons for the difference: 
an interaction of individual binaries with gravitational field of a cluster 
and a close fly-by (a close passages of a star near a binary system).

The first effect makes close binaries (so called ``hard'' systems) closer,
while wide systems become wider. 
The boundary between the two types of systems is determined 
by the condition $v_{\mathrm{orb}}\simeq v_{\mathrm{cluster}}$.
Here $v_{\mathrm{orb}}$ -- orbital velocity of a binary,
$v_{\mathrm{cluster}}$ -- mean velocity of stars in a cluster.

There are various effects of a close encounters. 
A  star flying-by can replace one of the binary component, 
can disrupt the binary, can induce a merging of the binary, 
or can just change binary orbital parameters. Probabilities of all of these
effects were studied in~\cite{Hut&Bahcall83,Heggie&Hut93}.
If a binary interact with another binary system, then, of course, there 
are more possible outputs of the interaction
(see~\cite{Rasio&95,Fregeau&04}).

Taking into account the possibility of a close fly-by, 
i.e pair stellar interaction, we destroy
one of the basic PS principle about an independence of 
objects' evolution. However, if
point-to-point interaction remain rare,  we still can use a PS in some
modified form.

There are two possible ways to modify the PS scheme to allow for close
encounters.
One method is to use a complete binary evolution scenario with 
simplified stellar dynamics. Another one, 
vice versa, includes precise N-body calculations of stellar motions with
a reduced evolutionary description. 
The second version was
realized in a recent series of papers by Portegies Zwart, Aarseth et
al.~\cite{Aarseth99,Hurley&01,Pooley&03,Gualandris&04,Fregeau&04}. 
Although this is just a first step in realization of a full scenario of
evolution in dense clusters.


\section{Population synthesis of isolated neutron stars}

 In this section we discuss a PS of isolated neutron stars
(NSs).\footnote{Russian-speaking readers can find more information on this 
subject in our review \cite{Popov&Prokhorov03}. It is also available on the Web 
at 
http://xray.sai.msu.ru/$\sim$polar/html/kniga.html.}
We focus on three types of research: radio pulsars studies, 
modeling of old NS evolution, 
and studies of close-by cooling NSs.  
The first two include as the main ingredients initial period, velocity, and
magnetic field distributions together with 
evolutionary laws for all these quantities.
In addition such a parameter as the angle between spin and magnetic axes can
be included. For the PS of cooling NSs ingredients are quiet
different, as the thermal history is not tightly connected with periods and
magnetic fields, instead the mass spectrum of NSs is important.

\subsection{Radiopulsars}

 As radio pulsars are the most known example of NSs and form the largest
observed population of these objects, it is not a big surprise that
there were several PS studies dedicated to these sources. 
The crucial feature of such models
is the necessity of simulating of the detectability of radio pulsars, i.e. it
is not enough to calculate an artificial population of NSs, 
it is also necessary to imitate a radio survey with realistic parameters
to compare calculations with observations.

 It is possible 
to distinguish three main reasons for a PS of radio
pulsars. The first one is related to a determination of initial
parameters of NSs (periods, magnetic fields, velocities). 
The second  deals with parameters of the evolution (for example the magnetic
field decay). 
And finally the third task is to predict properties of some unobserved
population (highly magnetized pulsars, gamma-ray pulsars, etc.)
basing on some known initial and evolutionary parameters.
   
One of the key questions concerning the evolution of
the population of radio pulsars is
the following: {\it does the magnetic field of NSs decay}?
Bhattacharya, Hartman, Verbunt and Wijers address this question in
a series of papers (see \cite{Verbunt&99} 
and references to earlier papers therein).
They use Monte-Carlo simulations to reproduce distributions of radio
pulsars on the $P-\dot P$ plane. Also the authors calculate velocity
distribution of NSs, their spatial distributions in different projections,
and dispersion measure distributions.   Verbunt et al. assume initial
magnetic field distribution of NSs similar to the one observed for radio
pulsars: a gaussian in a logarithmic scale. They also take into account
observability of the modeled sources. These authors conclude that the best fit
includes the magnetic   field decay on the time scale $\sim10^8$~yrs, and
that significant amount of NSs are born with velocities below
200~km~s$^{-1}$.

There are several hypothesis about initial conditions and 
evolutionary laws made by these authors that can be criticized.
Both velocity distributions used by the authors are  single-mode ones.
Nowadays the bimodality of the velocity distribution is considered to be a
standard assumption. Also modern observations strongly suggest that the
number of high velocity NSs is larger than it follows from the distributions
used by Verbunt et al.
The initial magnetic field distribution used by the authors hardly can
explain the existence of magnetars. 
Finally Verbunt et al. do not discuss the r\^ole of the inclination angle
$\chi$
(between spin and magnetic axes) and its evolution.

A more updated PS of radio pulsars is performed by Regimbau
and de Freitas Pacheco \cite{Regimbau&deFreitasPacheco01}. 
They include into the model
the evolution of $\chi$ and discuss the appearance of magnetars. 
A comparison with the papers by Bhattacharya et al.
shows how changes in parameters of a model of the PS change
the final results.  Regimbau and de Freitas Pacheco obtain an agreement
of the observed data with a model without the magnetic field decay, but
with an inclusion of an 
evolving (increasing) angle $\chi$. Another important conclusion
reached in \cite{Regimbau&deFreitasPacheco01} 
is related to magnetars. The authors obtain that about
23\% of NSs are highly magnetized ($B>10^{14}$~G). Their results suggest that
there is no bimodality in the magnetic field distribution, magnetars
naturally appear from the high field tail of the distribution which also
includes radio pulsars.

A separate topic of radio pulsar studies is related to $\gamma$-ray
observations. Gonthier et al. \cite{Gonthier&04} 
address the question of active
radio pulsar detection in the high-energy band. 
Such a research is very relevant now in the light of future space projects
as AGILE and GLAST. In the study by Gonthier et al. several 
new important parameters appear as the
authors incorporate independent models for the radio and gamma-ray beam
geometry.

 An unexpected result obtained by these authors is the necessity of the
magnetic field decay with relatively short time scale -- 2.8 Myr.
Predictions for AGILE and GLAST observations are very optimistic.
Gonthier et al. find that GLAST can observe up to 600 NSs in gamma-rays.

Beskin and Eliseeva \cite{Beskin&Eliseeva05} 
study a possibility of observation of 
dead pulsars as $\gamma$-ray sources. Strictly speaking these authors do not
perform a PS study, nevertheless their work can be considered
as an important input for population models.
They obtain a period distribution of dead (i.e. non-active) pulsars for two
models of the evolution of the angle $\chi$ between spin and magnetic axes.
Up to our knowledge a PS of NSs which are not active radio pulsars 
with a detailed model of the evolution of $\chi$ has never been performed.    
Such study can be very important as it can change conclusions about the
stage distribution of old NSs which we are going to discuss in the next
subsection.  


\subsection{Neutron star census and isolated accretors}

 Since early 70s different authors discuss a possibility of observations
of isolated accreting NSs. During these years conclusions changed from very
optimistic (nearly all old NSs are accreting; it is possible to observe
thousands of them with X-ray satellites like ROSAT) 
to very pessimistic (only a small
fraction of isolated NSs can reach the stage of accretion; most of accretors
are very dim sources).
 In 90s Monte-Carlo simulations of the NS population were performed by several
groups. We mention here two illustrative papers by Blaes and Rajagopal
\cite{Blaes&Rajagopal91} and Manning et al.~\cite{Manning&96}. 

The authors of \cite{Blaes&Rajagopal91}  
are motivated by a discussion on the nature of
gamma-ray bursts. Their PS study has a goal to calculate the distribution of 
masses accreted onto isolated NSs in the solar neighborhood.   
The main ingredients of the model are: initial spatial distribution of NSs,
their initial velocity distribution, the galactic potential, interstellar
medium (ISM) distribution, type of accretion. Blaes and Rajagopal (as many
other researchers before mid-90s) assume that all NSs relatively quickly
start to accrete. This assumption is the main mistake of many models at
that period, as the evolution of NSs has been absolutely ignored. Due to this
approach all groups obtained a huge number of accreting isolated NSs.

 We want to make one very important comment concerning the evolution of
isolated NSs. Before 1994 it was assumed that the velocity distribution of
NSs allows an existence of a big number of low-velocity compact objects.
After the paper by Lyne and Lorimer \cite{Lyne&Lorimer94} 
has appeared it became clear that
quite oppositely NSs (at least radio pulsars) are very high-velocity objects. 
The present day velocity
distribution \cite{Arzoumanian&02} 
is bimodal with more than 1/2 of NSs with velocities
larger than 500--600 km~s$^{-1}$. Very often different authors claim that 
small (actually zero) number of discovered isolated accretors 
can be explained by low luminosities due to high velocity of NSs
(in the Bondi formula the accretion rate
is inversely proportional to a cube of velocity). In some sense this
statement can be put in such a way: there are a lot of accretors, but with
modern satellites we cannot detect them.
However, it is not the case! 
NSs with big initial velocities just never reach the stage of accretion.  The 
number of isolated accretors is as low as the number of low velocity NSs. 

Manning et al. \cite{Manning&96} 
discuss the duration of stages preceding the 
onset of accretion. On the other hand,
they do not use the new (Lyne and Lorimer)
velocity distribution. That is why they find that a significant fraction of
NSs could reach the accretion stage in a lifetime of the Galaxy.  
Anyway, as Manning et al. included the evolution before accretion starts, 
probably their analysis  was the most complete till the end
of 90s.

 An attempt to model a global evolutionary distribution of isolated NSs 
is made in \cite{Popov&00a}. 
This {\it neutron star census} is performed to obtain
fractions of NS population on each of the four main evolutionary stages:
{\it ejectors, propellers, accretors, georotators} (see 
\cite{LPP96} for the stages
description). From the present day point of view this study has several
drawbacks. The authors use a very simple initial period and magnetic field
distributions. The velocity distribution is just a single-mode one.
The propeller and georotator stages are treated in a simplified way.
The r\^ole of $\chi$ is not studied. 
Nevertheless, in our opinion, 
the main features of the realistic distribution are picked up correctly.
For standard magnetic fields and without decay a NS spends most of its life
as an ejector (or as a georotator if velocity is high). For decaying fields
the propeller stage can be the most populated.

 In \cite{Popov&00a} it is 
found that for reasonable velocities more than 90\% of isolated NSs
spend all their lives as ejectors.
Fractions of propellers and georotators are found to be small.
However, we note that the authors treat
the georotator stage in a simplified manner, and depending on definitions
it is possible to say that significant fraction of very high-velocity NSs
are georotators. Also the stage of subsonic propeller 
(see \cite{Ikhsanov03} and
references therein) is not taken into account. 

A more detailed study of properties of accreting isolated NSs is performed
in \cite{Popov&00b}. The authors obtain the Log N -- Log S distribution for
accretors. It is shown that at low fluxes (say, below
$10^{-13}$-$10^{-14}$~erg~s$^{-1}$) accretors can become more abundant than
coolers (see the next subsection on that type of sources). Also velocity,
temperature and accretion rate distributions of isolated accreting NSs are
derived. 

To conclude this subsection we want to stress once again that PS models for
old isolated NSs passed a long way. Many parameters were significantly
refined. Nevertheless, as details of magneto-rotational evolution of NSs are
still unknown (evolution of $\chi$, properties of the propeller stage,
accretion rate and efficiency, etc.) 
in future more detailed calculations will inevitably appear.   


\subsection{Close-by young neutron stars}

 Quite a different branch of NS population studies is related to
young close-by cooling objects -- {\it coolers}. 
Presently, thanks to the ROSAT satellite, we know a set of 
nearly a dozen young hot
close-by NSs of different nature (see the list in \cite{Popov&03a}). 
Among these sources there are four normal radio pulsars, Geminga and a
geminga-like object, and seven radioquiet NSs (so-called ''Magnificent
seven''). Of course, this set of sources is not complete: there should be more
objects of that kind in the solar vicinity, it is just very difficult to
identify them in the crowded environment close to the Galactic plane.
Nevertheless, as all the sources were detected by one instrument (and if
possible we select them with upper bounds on age and distance from the Sun)
this set in uniform enough to perform a PS not worrying a lot about
selection effects (but see below on possible correlations between different
parameters of NSs).
The goal of such a study can be twofold: to understand the
origin of an observed  population of close-by {\it coolers}, and to test
the theoretical cooling curves by the observed objects.
The former approach was used in \cite{Popov&03a,Popov&03b}. 
The latter one -- in~\cite{Popov&04}.

 The main ingredients of such studies include:

\begin{itemize}
\item Initial spatial distribution of NSs (or, say, of progenitor stars); 
\item Initial velocity distribution of NSs;
\item Mass spectrum of NSs;
\item Cooling curves;
\item Properties of NS emission;
\item Properties of the ISM (to take absorption into account).
\end{itemize}

Some of the ingredients are known well enough for the purposes of  this
particular problem
(spatial and velocity distributions, and the ISM properties).
Others are poorly known.

In \cite{Popov&03a} 
it was shown that the local population of hot NSs from which
the thermal emission is detected is genetically connected with the Gould
Belt. This structure is responsible for the overabundance of massive stars
(and so of young compact objects) in the Solar vicinity. 
 
The authors of \cite{Popov&04} 
suggested to use Log N -- Log S distribution of
close-by coolers as the second additional test for calculations of the NS
thermal evolution. Indeed, if all but one ingredients of the PS are known,
then comparison with an actual data can be used to test the remaining
ingredient. 

Presently the only way to compare calculations of the thermal evolution
of NSs with observational data is the so-called T-t (temperature vs. age)
test. Being the most natural test for this problem it has some drawbacks:

\begin{itemize}
\item There are uncertainties in temperature and ages of observed sources;
\item The test is not sensitive enough for ages $>$few $10^5$ yrs;
\item The set of objects is not uniform at all.  
\end{itemize} 

Log N -- Log S as the second test has a complimentary to the T-t test
set of advantages and disadvantages. 

\begin{itemize}
\item There is no uncertainty in observational data;
\item The set of sources is nearly uniform;
\item The test is mainly sensitive to properties of NSs with
ages $>$few $10^5$ yrs.
\end{itemize}

On the other hand, there are some drawbacks (see \cite{Popov&04} 
for more details). 

Preliminary results show that the Log N -- Log S test works well as an addition
to the T-t test (we note, that the idea is to use the two tests together, not
Log N -- Log S alone). This is a good example how an
astrophysical PS study can be used to test predictions of physics of
matter in extreme states. 


\subsection{Unsolved problems}

Many ingredients of the PS models described above are not well known.

In our opinion the main difficulty of the radio pulsar PS is connected with
the fact that we do not understand well enough  emission mechanisms  in
different energy ranges.  Due to this it is difficult to derive many crucial
parameters and to produce results which can be directly compared with
observations. 
Many authors obtain drastically different results
even if all parameters of a model are the same, but 
mechanisms of radio pulsar emission are different.  
As a rule such parameters as
initial period and magnetic field distributions are not well
known. Evolutions of the magnetic field and of the angle between spin and
magnetic axes are unknown too.

A PS of old NSs suffers from the lack of observational data. No old 
isolated NS which is not  a millisecond radio pulsar is known, and there is
very little hope to find a way to observed an isolated NS at a stage of
propeller or georotator. 
Among theoretical uncertainties in addition to the ones described in the
previous paragraph it is necessary to mention difficulties with the
description of the propeller stage. Also the regime of accretion onto an
isolated NS and its efficiency is not known \cite{Popov&Treves2004}, 
estimates made with the Bondi
formula can be just an upper limit. 

Calculation of the number of observable coolers is the subject of
uncertainties in the mass spectrum of NSs, in cooling curves and in 
emission properties (spectrum and surface emissivity).


\section{Other examples of population synthesis}

\subsection{Stellar populations and spectral studies}

 The story of PS of stellar populations started in early 70s (see for
example \cite{Bruzual&Charlot03} 
for early references and for a brief historical introduction).
Historically the {\it empirical population synthesis} appeared earlier.
Nowadays both methods are actively used.
To learn more about the {\it evolutionary population synthesis} we can recommend
the paper \cite{Bruzual&Charlot03} 
and references therein. The {\it empirical population
synthesis}  is well described in \cite{CidFernandes&01}. 
Both methods together are
discussed in~\cite{Fritze-v.Alvensleben00}.

 Both methods are mainly motivated by the same problem: 
to derive properties of a
stellar content (ages, chemical composition, etc) basing on an integral
spectrum of a galaxy 
(an inverse problem, to predict a spectrum of a galaxy, is also possible).
The {\it empirical population synthesis} is based on a {\it decomposition} 
of a galactic spectrum to known spectra of stars of star populations.
I.e. it is necessary to find a linear combination of  known spectra which
fits an observed one (very often equivalent widths of spectral lines are
used as quantities to fit). 
The main problem of the method is connected with
non-uniqueness of a solution due to a number of free parameters. For
example,
there is a known degeneracy between age and metallicity: changes in
metallicity can mimic a change in the age of a population. An increased 
metallicity
and an increased age both make the integrated spectral energy distribution 
redder (see for example a discussion in~\cite{Zhang&04b}). 
       
 The main ingredients of the {\it evolutionary PS} are:

\begin{itemize}
\item Initial mass function;
\item Starformation rate;
\item Initial chemical composition and a rate of chemical enrichment.
\end{itemize} 

 Main problems are connected with non-sufficiently understood stages of the 
stellar
evolution -- stages after the main sequence 
(see for example~\cite{Maraston04}). 
Also there is a degeneracy
between age and initial mass function (it can be important for age
determinations of starbursts~\cite{Leitherer04}), for example, a smaller
fraction of massive stars can mimic an effect af aging of a stellar
population. 

 Both methods use some input spectral information.
In the case of the {\it evolutionary PS} it includes: spectra parametrized by
chemical composition, surface gravity and effective temperature.
For the {\it empirical PS} 
input data contains spectra of some types of stellar
populations, for example, spectra of clusters of stars (so-called {\it
single stellar populations}) with the same
chemical composition. Ages of stars belonging to each cluster are the same, 
but ages of different clusters are different, 
so one can fit an observed spectrum by a
composition of star clusters born at different time. 

 For years the PS of normal stars was done without taking into account the
r\^ole of binary stars. Only in late 90s several groups of authors started
to include binaries in their models (see \cite{Mas-Hesse&Cervino99} 
and references therein).
Among recent papers we want to mention \cite{Zhang&04} 
and a review in~\cite{DeDonder&Vanbeveren04}.

 All the authors who studied the influence of binary systems on the integral
spectral characteristics in the visual waveband conclude that its r\^ole
is very important. Especially it is important if Wolf-Rayet stars can form
only in binaries (an alternative view requires too strong stellar winds for
single stars).  

 Another specific problem is connected with galaxies with high rate of star
formation \cite{Leitherer04b}. 
Modeling young starbursts is a knotty problem due to several reasons: 

\begin{itemize}
\item Spectra are very complicated as 
still there are many bright massive stars.  
\item As stars were not born simultaneously
     stellar ages are uncertain and the level of this uncertainty is about  
      stars ages.
\item There is a lot of dust and the interstellar medium is very inhomogeneous.
\end{itemize}

An age determination for young star forming systems is discussed in some details 
in \cite{Leitherer04}. 
The approach is based on single stellar populations, with
addition of different (depending on time since beginning of a starburst) 
age indicators.  Depending on an age of a population different spectral lines 
(or other features) can be used as such indicators. For example, for 
very young stars ($\sim$4 Myr) 
nebular emission lines  and UV stellar-wind lines can be used
as age discriminants. 

An example of determination of an age of an old system and references for early
papers are given in~\cite{Zhang&04b}. 
These authors also develop a model based on
single stellar populations for ages 1-19 Gyr. Applying the model to the
galaxy M32 (a satellite of the Andromeda galaxy) they obtain that it can
be described as a single population with a solar metallicity and an age
$\sim$6.5 Gyr.

Evolutionary synthesis was used in \cite{Bell&deJong01} 
to obtain such an important
characteristic as mass to light (M/L) ratio for galaxies.
This quantity is widely used to estimate a mass of a galaxy knowing its
luminosity in some band.


\subsection{Active galactic nuclei and X-ray background}

Observations aboard ROSAT and, especially, Chandra 
satellites have shown that more
than 60\% (up to 90\% in softer X-rays) of the X-ray background (XRB) 
could be
resolved, i.e. point sources are responsible for that radiation.
According to optical identifications most of these objects 
are active galactic nuclei (AGNs) with small addition of clusters of
galaxies and stars.  
Such observations rised interest to PS studies of AGNs. 

Initially PS studies of the XRB were motivated by the so-called unified model
of AGN (see \cite{Pompilio&00,Madau&94} for examples of PS calculations 
and for references to earlier papers). 
In this model all AGN activity is explained by
accretion onto central supermassive black hole (see \cite{Antonucci93} 
for the unified model description). 
Different types of AGNs
appear due to different black hole masses, different properties of the
accreting medium and, especially, due to obscuration by a torus surrounding
the black hole.  Later on PS models became more phenomenological as all main
ingredients are taken from observational data 
(for example~\cite{Gilli&01,Gandhi&Fabian03,Halevin03}).

Among the ingredients we can mention:

\begin{itemize}
\item Relative fraction of obscured and non-obscured AGNs;
\item Luminosity functions of AGNs;
\item Spectral energy distribution;
\item Evolution of these parameters (if any).
\end{itemize}

Usually models are closer to those called ``population synthesis'' by
Fritze - v. Alvensleben \cite{Fritze-v.Alvensleben00}
than to the scenarios described in sec.~4--5
(which must be called "evolutionary synthesis" in her notation).
In studies of the XRB authors do not follow the evolution of AGNs starting
with some initial conditions 
(seed black holes in centers of galaxies, or even from the scenario of
hierarchical clustering).
In most of papers the XRB is explained by contribution of absorbed (type 1) and
non-absorbed (type 2) AGNs.  Spectra of type 2 AGNs can be calculated from
spectra of the type 1 by adding some obscuration. This is the main
contribution of the unified model to the PS scenario. In early papers the
ratio of numbers of type1 to type2 sources was also taken as the unified model 
prediction, 
but now it is taken from the observational data. 

Nowadays these models can more or less successfully explain the main properties 
of
the XRB. Nevertheless, it is necessary to make them less phenomenological by
linking the PS calculations with calculations of the early evolution of the
galaxies (including galaxy formation, seed black holes, etc.). Also the
r\^ole of starburst activity in the XRB is not known well enough 
(see~\cite{Halevin03}).


\section{Conclusions}

 In this lecture notes we briefly described the main applications of 
{\it population synthesis} in astrophysics. The spectrum of applications is very
wide: from normal stars to compact remnants, from integral spectra of
galaxies to X-ray background. We mainly focused on evolution of isolated and
binary compact objects as these subjects are closer to our personal
scientific interests.

 The population synthesis as a method of theoretical (numerical) studies
is a typically astrophysical tool as in astronomy
very often we can observe just a tiny
fraction of a population of sources, or cannot resolve single sources at all.  
In these cases the population synthesis can help to derive or check
parameters of the whole set of studied objects, or to predict properties of
an unobserved fraction of the sources.

 As our knowledge about some population of objects grows models of the
population synthesis become more complicated. Finally, when
the whole population can be studied observationally, and also an evolution of 
its
members can be directly derived from actual data, population modeling
becomes unnecessary (except to show self-consistency of results of
observations). However, in most of astrophysical situations we are very far
from this ideal, and so the population synthesis calculations will be an
important tool for a long time.  

\section*{Acknowledgments}

We want to thank the Organizers of the School
for financial support and hospitality.
S.P. acknowleges a postdoctoral fellowship from the University of Padova
where he worked at the time of the School.
This work was partly supported by RFBR grants 04-02-16720 and 03-02-16068.


\end{document}